\documentclass{LMCS}
\usepackage{amssymb}
\usepackage{amsmath}
\usepackage{amsfonts}
\usepackage{hyperref}

 \def\pushright#1{{
    \parfillskip=0pt            
    \widowpenalty=10000         
    \displaywidowpenalty=10000  
    \finalhyphendemerits=0      
    \leavevmode                 
    \unskip                     
    \nobreak                    
    \hfil                       
    \penalty50                  
    \hskip.2em                  
    \null                       
    \hfill                      
    {#1}                        
    \par}}                      

\newcommand{\ens}[1]{\{ #1\}}

\newcommand{\cA}{{\mathcal A}}

\newcommand{\ourcomment}[1]{}

\newcommand{\cL}{{\mathcal L}}
\newcommand{\cG}{{\mathcal G}}

\newcommand{\cU}{{\mathcal U}}

\newcommand{\met}{\mathcal{M}}
\newcommand{\pmu}{\mathcal{PMU}}

\newtheorem{theorem}[thm]{Theorem}
\newtheorem{corollary}[thm]{Corollary}
\newtheorem{lemma}[thm]{Lemma}

\newtheorem{definition}[thm]{Definition}
\newtheorem{example}[thm]{Example}

 \def\pushright#1{{
    \parfillskip=0pt            
    \widowpenalty=10000         
    \displaywidowpenalty=10000  
    \finalhyphendemerits=0      
    \leavevmode                 
    \unskip                     
    \nobreak                    
    \hfil                       
    \penalty50                  
    \hskip.2em                  
    \null                       
    \hfill                      
    {#1}                        
    \par}}                      

\newcommand{\fullpaper}[1]{#1}
\newcommand{\abstractonly}[1]{}

\newcommand{\next}{{\tt next}}
\newcommand{\cS}{{\mathcal S}}
\newcommand{\traces}{{\tt Traces}}
\newcommand{\genstate}[2]{\langle #1,\mathbf{\vec{#2}} \rangle}
\newcommand{\rate}[1]{\mathbf{\vec{#1}}}
\newcommand{\clock}[1]{\mathbf{\vec{#1}}}
\newcommand{\cadlag}[1]{{\mathcal D}^{#1}[0, \infty)}
\newcommand{\Gr}[1]{{\tt Graph}(#1)}
\newcommand{\delay}[1]{{\tt delay}_{#1}}
\newcommand{\gs}{{\tt g}_{\tt s}}
\newcommand{\genst}[1]{\mathtt{g}_{#1}}
\newcommand{\cF}{{\mathcal F}_k}
\newcommand{\timeleft}{{\mathcal T}}

\def\doi{2 (1:4) 2006}
\lmcsheading%
{\doi}
{23}
{}
{}
{Nov.~\phantom{0}5, 2004}
{Mar.~\phantom{0}7, 2006}
{}

\begin{document}
\title{Approximate reasoning for real-time probabilistic processes}

\author[V.~Gupta]{Vineet Gupta\rsuper a}
\address{{\lsuper a}Google Inc.}
\email{vineet@google.com}
\author[R.~Jagadeesan]{Radha Jagadeesan\rsuper b}
\address{{\lsuper b}School of CTI, DePaul
University} \email{rjagadeesan@cs.depaul.edu}
\thanks{{\lsuper b}Research supported in part by NSF 0244901 and NSF 0430175.}

\author[P.~Panangaden]{Prakash Panangaden\rsuper c}
\address{{\lsuper c}School of Computer Science, McGill University}
\email{prakash@cs.mcgill.ca}
\thanks{{\lsuper c}Research supported in part by NSERC and EPSRC}

\begin{abstract}

  We develop a pseudo-metric analogue of bisimulation for generalized
  semi-Markov processes.  The kernel of this pseudo-metric corresponds to
  bisimulation; thus we have extended bisimulation for continuous-time
  probabilistic processes to a much broader class of distributions than
  exponential distributions.  This pseudo-metric gives a useful handle on
  approximate reasoning in the presence of numerical information --- such
  as probabilities and time --- in the model.

  We give a fixed point characterization of the pseudo-metric.  This makes
  available coinductive reasoning principles for reasoning about distances.
  We demonstrate that our approach is insensitive to potentially ad hoc
  articulations of distance by showing that it is intrinsic to an
  underlying \emph{uniformity}.  We provide a logical characterization of
  this uniformity using a real-valued modal logic.

  We show that several quantitative properties of interest are continuous
  with respect to the pseudo-metric.  Thus, if two processes are metrically
  close, then observable quantitative properties of interest are indeed
  close.
\end{abstract}

\maketitle

\vskip-\bigskipamount
\section{Introduction}

The starting point and conceptual basis for classical
investigations in concurrency are the notions of equivalence and
congruence of processes --- when can two processes be considered
the same and when can they be substituted for each other?   Most
investigations into timed~\cite{Alur92,Alur94} and probabilistic
concurrent processes are based on equivalences of one kind or
another,
e.g.~\cite{Cleaveland92,Hansson94,Hillston94,Larsen91,Segala95,Philippou00}
to name but a few.

As has been argued before~\cite{Jou90,Desharnais99b,Desharnais04},
this style of reasoning is fragile in the sense of being too
dependent on the exact numerical values of times and
probabilities.  Previously this had pointed out for probability,
but the same remarks apply, \emph{mutis mutandis}, to real time as
well. Consider the following two paradigmatic examples:
\begin{itemize}
\item Consider the probabilistic choice operator: $\cA_1 +_p
\cA_2$, which starts $\cA_1$ with probability $p$ and $\cA_2$ with
probability $1-p$.  Consider $\cA_1 +_{p +\epsilon} \cA_2$ and
$\cA_1 +_{p +2\epsilon} \cA_2$.  In traditional exact reasoning,
the best that one can do is to say that all these three processes
are inequivalent.  Clearly, there is a gradation here: $\cA_1 +_{p
+\epsilon} \cA_2$ is closer to $\cA_1 +_{p} \cA_2$ than $\cA_1
+_{p +2 \epsilon} \cA_2$ is to $\cA_1 +_{p} \cA_2$.

\item Consider the ${\tt delay}_t.\cA$ operator that starts $\cA$
after a delay of $t$ time units.  Consider ${\tt delay}_{t+
\epsilon}.\cA$ and ${\tt delay}_{t+2\epsilon}.\cA$.  Again, in
exact reasoning, the best that one can do is to say that all these
three processes are inequivalent.  Again, ${\tt delay}_{t+
\epsilon}.\cA$ is intuitively closer to ${\tt delay}_t.\cA$ than
${\tt delay}_{t +2 \epsilon}.\cA$ is to ${\tt delay}_t.\cA$.
\end{itemize}
In both examples, the intuitive reasoning behind relative
distances is supported by calculated numerical values of
quantitative observables --- expectations in the probabilistic
case and (cumulative) rewards in the timed case.

The fragility of exact equivalence is particularly unfortunate for
two reasons: firstly, the timings and probabilities appearing in
models should be viewed as numbers with some error estimate.
Secondly, probability distributions over uncountably many states
arise in even superficially discrete paradigms such as Generalized
semi-Markov processes (e.g. see~\cite{Shedler87} for a textbook
survey), and discrete approximations are used for algorithmic
purposes~\cite{Baier99,Haverkort02}.  These approximants do not
match the continuous state model exactly and force us to think
about approximate reasoning principles --- e.g. when does it
suffice to prove a property about an approximant?

Thus, we really want an ``approximate'' notion of equality of
processes.  In the probabilistic context, Jou and
Smolka~\cite{Jou90} propose that the correct formulation of the
``nearness'' notion is via a metric.  Similar reasons motivate the
study of Lincoln, Mitchell, Mitchell and Scedrov~\cite{Lincoln98},
our previous study of metrics for labelled Markov
processes \cite{Desharnais99b,Desharnais04,Desharnais02a}, the
study of the fine structure of these metrics by van Breugel and
Worrell~\cite{vanBreugel01a,vanBreugel01b} and the study of
Alfaro, Henzinger and Majumdar of metrics for probabilistic
games~\cite{dealfaro03}.

In contrast to these papers, in the present paper we focus on real-time
probabilistic systems that combine \emph{continuous time and probability}.
We consider generalized semi-Markov processes (GSMPs).  Semi-Markov
processes strictly generalize continuous-time Markov chains by permitting
general (i.e.\ non-exponential) probability distributions; GSMPs further
generalize them by allowing competition between multiple events, each driven
by a different clock.

Following the format of the usual definition of bisimulation as a maximum
fixed point, we define a metric on configurations of a GSMP as a maximum
fixed point.  This permits us to use analogues of traditional coinductive
methods to reason about metric distances.  For example, in exact reasoning,
to deduce that two states are equivalent, it suffices to produce a
bisimulation that relates the states.  In our setting, to show that the
distance between two states is less than $\epsilon$, it suffices to produce
a (metric) bisimulation that sets the distance between the states to be
less than $\epsilon$.

Viewing metric distance $0$ as bisimilarity, we get a definition of
bisimulation for GSMPs, a class that properly includes CTMCs.  In contrast
to existing work on bisimulation for general probability distributions
(e.g.~\cite{Bravetti02,Hermanns02}) our definition accounts explicitly for
the change of probability densities over time.

Secondly, we demonstrate that our study does not rely on any ``ad-hoc''
construction of metric distances.  Uniform spaces capture the essential
aspects of metric distances by axiomatizing the structure needed to capture
relative distances -- e.g.  statements of the form ``x is closer to y than
to z.''  A metric determines a uniform space but different metrics can
yield the same uniform space.  Uniform spaces represent more information
than topological spaces but less than metric spaces, so we are identifying,
as precisely as we can, the intrinsic meaning of the quantitative
information.  We present our maximal fixpoint construction as a
construction on uniform spaces, showing that the numerical values of
different metric representations are not used in an essential way.  In
particular, in our setting, it shows that the actual numerical values of
the discount factors used in the definition of the metric do not play any
essential role.

Thirdly, we provide a ``logical'' characterization of the uniformity using
a real-valued modal logic.  In analogy to traditional completeness results,
we prove that the uniformity notion induced by the real-valued modal logic
coincides with the uniformity induced by the metric defined earlier.  Our
logic is intentionally chosen to prove this completeness result.  It is not
intended to be used as an expressive specification formalism to describe
properties of interest.  Our framework provides an intrinsic
characterization of the quantitative observables that can be accommodated
-- functions that are continuous with respect to the metric.

Finally, we illustrate the use of such studies in reasoning by showing that
several quantitative properties of interest are continuous with respect to
the metric.  Thus, if two processes are close in the metric then observable
quantitative properties of interest are indeed close.  For expository
purposes, the list considered in this paper includes expected hitting time,
expected (cumulative and average) rewards.  The tools used to establish
these results are ``continuous mapping theorems'' from stochastic process
theory, and provide a general recipe to tackle other observables of
interest.

The rest of this paper is organized as follows.  We begin with a review of
the model of GSMPs in Section~\ref{sec4}.  We then give a review of the
basic ideas from stochastic process theory -- metrics on probability
measures on metric spaces in Section~\ref{sec2} and the Skorohod J2 metrics
on timed traces in Section~\ref{sec3}.  We discuss timed traces in the
context of GSMPs in Section~\ref{sec-cadl}.  We define metric bisimulation
in Section~\ref{sec5}.  We discuss interesting quantitative observables are
continuous functions in Section~\ref{sec7}.  We present our construction in
terms of uniform spaces in Section~\ref{sec8}.  Finally, we show the
completeness of the real-valued modal logic in Section~\ref{sec6}.

\section{Generalized semi-Markov processes}\label{sec4}
GSMPs properly include finite state CTMCs while also permitting
general probability distributions.  We describe GSMPs informally
here following the formal description of~\cite{Whitt80}.  The key
point is that in each state there are possibly several events that
can be executed.  Each event has its own clock - running down at
its own rate - and when the first one reaches zero that event is
selected for execution.  Then a probabilistic transition
determines the final state and any new clocks are set according to
given probability distributions: defined by conditional density
functions.  The probability distribution over the next states
depends only on the current state and the event that has occurred:
this is the ``Markov'' in semi-Markov.  The clocks are reset
according to an arbitrary distribution, not necessarily an
exponential (memoryless) distribution: hence the ``semi''.  We
will consider finite-state systems throughout this paper.

A finite-state GSMP over a set of atomic propositions $AP$ has the
following ingredients:
\begin{enumerate}
\item A finite set $S$ of states.  Each state has an associated
finite set of events $I(s)$, each with its own clock (we use the
same letter for an event and its clock) and a non-zero rate for
each clock in $I(s)$.  A clock in $I(s)$ runs down at the constant
rate associated with it.
\item A labelling function $\texttt{Props}: S \rightarrow 2^{AP}$ that
assigns truth values to atomic propositions in each state.
\item A continuous probability density function $f(t;s,i;s',i')$, over
  time, for each $i \in I(s)$, for each target state $s'$ and $i' \in
  I(s')$.  This is used to define how clocks are reset during a
  transition.
\item For each $i \in I(s)$, a probabilistic transition function $\next_i: S
\times S \rightarrow [0,1]$.  We require $\sum_{s' \in S}
\next_i(s,s') = 1$.
\end{enumerate}

We use $\clock{c}, \clock{c'}$ (resp. $\rate{r}$) for vectors of
clock values (resp. rates). We use the vector operation $\clock{c}
- \rate{r_c} t$ to indicate the clock vector resulting from
evolution of each clock under its rate  for time $t$.
\begin{definition}
Let $s$ be a state.  A \textbf{generalized state} is of the form
$\genstate{s}{c}$ where $\clock{c}$ is a vector of clock values
indexed by $i \in I(s)$ that satisfies a uniqueness condition:
there is a unique clock in $I(s)$ that reaches $0$ first.
\end{definition}
We write $\timeleft(\genstate{s}{c})$ for the time required for
the first clock (unique by the above definition) to reach $0$.  We
use  $\cG$ for the set of generalized states, and $\gs, \gs',\gs^1
\ldots $ for generalized states.

We describe the evolution starting in a generalized state
$\genstate{s}{c}$.  Each clock in $\clock{c}$ decreases at its
associated rate.  By the uniqueness condition on generalized
states, a unique clock reaches $0$ first.  Let this clock be $i
\in I(s)$. The distribution on the next states is determined by
the probabilistic transition function $\next_i: S \times S
\rightarrow [0,1]$.  For each target state $s'$,
\begin{itemize}
\item The clocks $i' \in I(s) \setminus I(s')$
are discarded.
\item The new clocks $i' \in I(s') \setminus [ I(s) \setminus \{ i \}
]$, get new initial time values assigned according to the
continuous probability density function $f(t;s,i;s',i')$.
\item The remaining clocks in $I(s) \cap I(s')$ carry forward their time values from $s$
to $s'$
\end{itemize}
The continuity condition on probability distributions ensures that
this informal description yields a legitimate Markov
kernel~\cite{Whitt80}.   The semantics of a real-time
probabilistic process can be described as a discrete-time Markov
process on generalized states.  For each generalized state, we
associate a set of sequences of generalized states that arise
following the prescription of the evolution given above.

\section{Pseudometrics}\label{sec2}
\begin{definition}
A pseudometric $m$ on a state space $\cS$ is a function $\cS
\times \cS \rightarrow [0,1]$ such that:
\[
m(x,x) = 0,   \ m(x,y) = m(y,x), \ m(x,z) \leq m(x,y) + m(x,z)
\]
\end{definition}
A function $f: (M,m) \rightarrow (M',m')$ is Lipschitz if $
(\forall x,y) \ m'(f(x),f(y)) \leq m(x, y)$.

We consider a partial order on pseudometrics on a fixed set of
states $\cS$.
\begin{definition}
$\met$ is the class of pseudometrics on $\cS$ ordered as: \[m_1
\preceq m_2 \ \mbox{ if } (\forall s,t)\ m_1(s,t) \geq m_2(s,t).\]
\end{definition}

\begin{lemma}
$( \met,\preceq)$  is a complete lattice.
\end{lemma}
The top element $\top$ is the constant $0$ function, and the
bottom element is the discrete metric~\cite{Desharnais02a}. Thus,
any monotone function $F$ on $( \met,\preceq)$ has a complete
lattice of fixed points.

\subsection{Wasserstein metric}\label{sec3a}
The Wasserstein metric is actually a prescription for lifting the
metric from a given (pseudo)metric space to the space of
probability distributions on the given space.

Let $(M,m)$ be a pseudometric space, and let $P,Q$ be probability
measures on $M$.  Then, $W(m)(P,Q)$ is defined by the solution to
the following linear program ($h : M \rightarrow [0,1]$ is any
function):
\begin{eqnarray*}
W(m)(P,Q) &=&\sup_h \int h dP - \int h dQ\\
&&\mbox{subject to}:
                    \forall s \in M.\ 0 \leq h(s) \leq 1 \\
                     &&\forall s,s'. \  |h(s) -  h(s')| \leq  m(s, s').
 \end{eqnarray*}

 An easy calculation using the linear program shows that the distances on
 distributions satisfies symmetry and the triangle inequality, so we get a
 pseudometric - written $W(m)$ - on distributions.

By standard results (see Anderson and Nash~\cite{Anderson87}),
this is equivalent to defining $W(m)(P,Q)$ as the solution to the
following dual linear program (here $\rho$ is any measure on $M
\times M$, $S$ and $S'$ are any measurable subsets):

\begin{eqnarray*}
W(m)(P,Q) &=&\inf_\rho  \int m\ d \rho \;\;\\
&&\mbox{subject to}: \begin{array}[t]{l}
                     \forall S.  \rho(S\times M) = P(S)  \\
                     \forall S'.  \rho(M \times S') = Q(S') \\
                     \forall S, S'.\ \rho(S\times S') \geq  0.
                     \end{array}
\end{eqnarray*}
The Wasserstein construction is monotone on the lattice of
pseudometrics.
\begin{lemma}\label{wass-mon}
$m  \preceq m' \Rightarrow W(m) \preceq W(m')$
\end{lemma}\fullpaper{
\proof{} Clearly every solution to the linear program for
$W(m')(P,Q)$ is also a solution to the linear program for
$W(m)(P,Q)$.   The result is now immediate. \qed}

We discuss some concrete examples to illustrate the distances
yielded by this construction.  Let $(M,m)$ be a $1$-bounded metric
space, i.e. $m(x,y) \leq 1$, for all $x,y$.  Let $1_x$ be the unit
measure concentrated at $x$,
\begin{example}\label{shift-lp}
We calculate $W(m)(1_x,1_{x'})$. The primal linear program --- using
the function $h: M \rightarrow [0,1]$ defined by $h(y) = m(x,y)$ ---
yields $W(m)(1_x,1_{x'}) \geq m(x,x')$. The dual linear program ---
using the product measure --- yields $W(m)(1_x,1_{x'}) \leq
m(x,x')$.
\end{example}
\begin{example}
Let $P,Q$ be such that for all measurable $U$, $|P(U) - Q(U)| <
\epsilon$.  For any $1$-bounded function $h$, $ |\int h dP - \int
h dQ| < \epsilon$ --- since for any simple function $g$ with
finite range $\{ v_1 \ldots v_n \}$ dominated by $h$:
\[
\begin{array}{l}
\sum_i v_i P(g^{-1}(v_i)) - v_i Q(g^{-1}(v_i)) \\
= \sum_i v_i [P(g^{-1}(v_i)) - Q(g^{-1}(v_i))]  \\
\leq \sum_{i=1}^k v_i [P(g^{-1}(v_i)) - Q(g^{-1}(v_i))] \\  
\phantom{\leq}\mbox{ (wlog assume } P(g^{-1}(v_i)) \geq Q(g^{-1}(v_i) \mbox{ for exactly } v_1,\ldots,v_k)\\
\leq \sum_{i=0}^k  P(g^{-1}(v_i)) - Q(g^{-1}(v_i)) \\
= P(g^{-1}(\{ v_1 \ldots v_k \})  - Q(g^{-1}(\{ v_1 \ldots v_k\})) \\
< \epsilon
\end{array}
\]
So,  $W(m)(P,Q) < \epsilon$.
\end{example}
In $1$-bounded metric spaces, the Wasserstein metric is closely
related to the Prohorov metric, $\pi$, which metrizes the topology
of weak convergence.  We say that $ P_n$ weakly converges to $P$
if for all bounded continuous real-valued functions $f$ $\int f
dP_n$ converges to $\int f dP$.  For any Borel set $A$, we write
$A^{\epsilon}$ for $\ens{u:\exists v\in A. d(u,v)<\epsilon}$.  The
Prohorov metric $\pi(P,Q)$ between two measures is defined by
\[ \inf_{\epsilon\geq 0} P(A)\leq Q(A^{\epsilon})+\epsilon \text{ and }
Q(A) \leq P(A^{\epsilon})+\epsilon.\] The connection between the
Wasserstein metric and the Prohorov metric, see~\cite{Gibbs01} for
a tutorial presentation of various such relationships, is:
\[
\sqrt{\pi(P,Q)} \leq W(m)(P,Q) \leq 2 \pi(P,Q).
\]

The following lemma is the key tool to approximate continuous
probability distributions by discrete distributions in a separable
metric space.  We use the following lemma later with the $U_i,
i>1$'s being subsets of $\epsilon$-neighborhoods of a point, and
the $U_i, i>1$ being a finite cover, wrt $P$,  for all but
$\epsilon$ of the space (which will be covered by $U_0$).
\begin{lemma}\label{random1}
Let $P$ be a probability measure on $(M,m)$.  Let $U_i, i
=0,1,2,\ldots n $ be a finite partition of the points of $M$ into
measurable sets such that:
\begin{itemize}
\item $(\forall\ i \geq 1) \ [ {\tt diameter} (U_i) \leq
\epsilon]$ \footnote{The diameter of a set $S$ is $\sup \{ m(x,y)
\mid x,y \in S \}$}
\item $P(U_0) \leq \epsilon$
\end{itemize}
Let $x_i$ be such that $x_i \in U_i$.  Define a discrete
probability measure $Q$ on $(M,m)$ by: $Q(\{ x_i \}) = P(U_i)$.
Then:
\[ W(m)(P,Q) \leq 2 \epsilon \]
\end{lemma}
\fullpaper{ \proof{} Let $h : M \rightarrow [0,1]$ be any function
that satisfies $\forall s,s'. \  |h(s) -  h(s')| \leq  m(s, s')$.
Then

\[
\begin{array}{rcl}
    \int h dP - \int h dQ &=& \sum_i \int_{U_i} h dP - \int_{U_i} h dQ\\
                          &=& \sum_i \int_{U_i} h dP - h(x_i)P(U_i)\\
                          &=& \sum_i \int_{U_i} (h(x) - h(x_i)) dP\\
                          &\leq& \int_{U_0} 1 dP + \sum_{i>0} \int_{U_i}
                          \epsilon dP\\
                          &\leq& P(U_0) + \int \epsilon dP\\
                          &\leq& 2\epsilon\\
\end{array}
\]

The fourth inequality follows as $0 \leq h(x) \leq 1$ and from our
assumption on $U_i, m(x, x_i) \leq \epsilon$, and from the
constraint on $h$, $h(x) - h(x_i) \leq m(x, x_i) \leq \epsilon$.
Thus $W(m)(P,Q) \leq 2\epsilon$.
%
%
\qed}

\section{Cadlag functions} \label{sec3}
Usually when one defines bisimulation one requires that ``each
step'' of one process matches ``each step'' of the other process.
What varies from situation to situation is the notion of step and
of matching.  In the present case there is no notion of atomic
step: one has to match sequences instead.  In the usual cases
matching steps and matching sequences are equivalent so one works
with steps as they are simpler.  Here we have no choice: we have
to work with timed sequences.

The timed sequences that one works with are functions from
$[0,\infty)$ to the state space.  Since we have discrete
transitions these functions are not continuous.  It turns out that
the class of functions most often used are the
\emph{cadlag}\footnote{This is an acronym for the French phrase
  ``continue \`a droite limites \`a gauche'' meaning ``continuous on the
  right with left limits.''} functions.
\begin{definition}
Let $(M,m)$ be a pseudometric space. $f: [0, \infty) \rightarrow
M$ is \textbf{cadlag} if for any decreasing sequence $ \{ t \}
\downarrow t_0 $
\[  \lim_{t \rightarrow t_0} f(t) = f(t_0) \]
and for any increasing sequence $\{ t \} \uparrow t_0$
\[ \lim_{t \rightarrow t_0} f(t) \mbox{ exists } \]
We write $\cadlag{(M,m)}$ (or $\cadlag{m}$, when $M$ is clear from
context) for cadlag functions with range $(M,m)$.
\end{definition}
These functions have very nice properties: for example, they have
at most countably many discontinuities.  More to the point
perhaps, if one fixes an $\epsilon > 0$, then in any bounded interval
there are at most finitely
many jumps higher than $\epsilon$, so all but finitely many jumps
are small.

The study of metrics on spaces of these functions was initiated by
Skorohod~\cite{Skorohod56}; see Whitt's book~\cite{Whitt02} for an
expository presentation.  Skorohod defined several metrics: we use
one called the $J_2$ metric.  The most naive metric that one can
define is the sup metric.  This fails to capture the convergence
properties that one wants: it insists on comparing two functions
at the exact same points. Skorohod's first metric (the $J_1$
metric) allows one to perturb the time axis so that functions
which have nearby values \emph{at nearby points} are close.  The
$J_1$ metric also fails to satisfy certain convergence properties
and we use the $J_2$ metric defined below, which like the $J_1$
metric, allows one to compare nearby time points.

Let $(M,m)$ be a metric space.  Let $|\cdot|$ be the metric on
positive reals $R^+$ be defined by
\[ |\cdot|(r,r') = |r-r'| \]
\begin{definition}[Skorohod $J_2$ metric] \hfill \\
Let $(M,m)$ be a metric space.  Let $f,g$ be cadlag functions:
$[0, \infty) \rightarrow M$.  $J(m)(f,g)$ is defined as:

\begin{eqnarray*}
\max ( && \sup_t \inf_{t'} [\max(m(f(t),g(t')), |t-t'|)], \\
        &&   \sup_{t'} \inf_{t} [\max(m(f(t),g(t')), |t-t'|)] \ \ \ ) \\
\end{eqnarray*}
\end{definition}

Thus the $J_2$ distance between two functions is the Hausdorff
distance between their graphs (ie. the set of points $(x,f(x))$)
in the space $[0,\infty) \times M$ equipped with the metric
$d((x,s), (y,t)) = max(|x-y|, m(s,t))$.

The following lemma is immediate from definitions.
\begin{lemma}\label{sk-mon}
$m_1 \preceq m_2 \Rightarrow J(m_1) \preceq J(m_2)$
\end{lemma}
The next lemma is standard, e.g. see
Billingsley~\cite{Billingsley99}.
\begin{lemma}[Skorohod] \label{cadlag-sep} \hfill \\
If $(M,m)$ is separable, $\cadlag{M}$ is a separable space with a
countable basis given by piecewise constant functions with
finitely many discontinuities and finite range contained in a
basis of $M$.
\end{lemma}

We consider a few examples, to illustrate the metric --- see the
book by Whitt~\cite{Whitt02} for a detailed analysis of Skorohod's
metrics.  The first example shows that jumps/disconti\-nuities can
be matched by nearby jumps.
\begin{example}\label{oneJ}~\cite{Whitt02}
Let $\{ b_n \}$ be an increasing sequence that converges to
$\frac{1}{2}$.  Consider:

\[ f_{b_n}(r) = \left\{
                \begin{array}{l}
                0, r < b_n \\
                1, r \geq b_n
                \end{array}
                \right.
\]
These are depicted in the picture below.

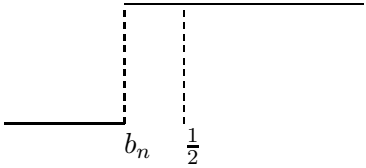
\setlength{\unitlength}{0.00625in}
\begin{picture}(300,200)(0,30)
\put(0,100){\line(1,0){100}}
\multiput(100,100)(0,10){10}{\line(0,1){5}}
\multiput(150,100)(0,10){10}{\line(0,1){5}}
\put(100,200){\line(1,0){200}}
 \put(150,75){$\frac{1}{2}$}
 \put(100,75){$b_n$}
\end{picture}

The sequence $\{ f_{b_n} \}$ converges to $f_{\frac{1}{2}}$.

\end{example}
The next example shows that a single jump can be matched by
multiple nearby jumps.
\begin{example}\label{twoJ}~\cite{Whitt02}
Let $\{ a_n, \}, \{c_n \}$ be increasing sequences that converges
to $\frac{1}{2}$ such that $a_n < c_n $.  Let:

\[ g_{n}(r) = \left\{
                \begin{array}{ll}
                0, &r < a_n \\
                1, &a_n \leq r < c_n \\
                0, &c_n \leq r < \frac{1}{2}\\
                1, &r \geq \frac{1}{2}
                \end{array}
                \right.
\]
These are depicted in the picture below.

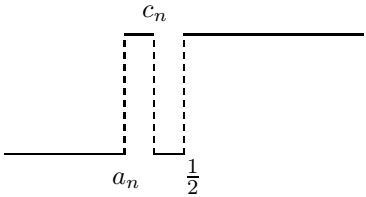
\setlength{\unitlength}{0.00625in}
\begin{picture}(300,200)(0,30)
\put(0,100){\line(1,0){100}}
 \put(125,100){\line(1,0){25}}
 \put(100,200){\line(1,0){25}}
\multiput(100,100)(0,10){10}{\line(0,1){5}}
\multiput(150,100)(0,10){10}{\line(0,1){5}}
\multiput(125,100)(0,10){10}{\line(0,1){5}}
\put(150,200){\line(1,0){150}}
 \put(150,75){$\frac{1}{2}$}
 \put(90,75){{\small $a_n$}}
 \put(115,215){{\small $c_n$}}
\end{picture}

The sequence $\{ g_{n} \}$ converges to $f_{\frac{1}{2}}$.

\end{example}
The next two non-examples shows that ``jumps are detected''.  In a
later section, we develop a real-valued modal logic that captures
the reasoning behind these two non-examples.  Here, to provide
preliminary intuitions, we provide a preview of this development
in a specialized form. Given a cadlag function $f$ with range
$[0,1]$ and the standard metric, and a Lipschitz function $h$ on
$[0,1]$ let $\cL(h)(f)$ be defined as
\[ \cL(h)(f)(t) = \sup_{t'} \{ h(f(t')) - | t' -t| \mid t' \in [0, \infty) \}
\]
In this definition, view $h$ as a test performed on the values
taken by the function $f$.  Since $h$ is a Lipschitz function on
$[0,1]$, the results of such tests are smoothed out, and include
the analogue of (logical) negation via the operation $1 - (\cdot)$
and smoothed conditionals via $h_q(x) = max(0, x-q)$ that
correspond to a ``greater than $q$'' test.   The $\cL(h)(f)$ also
performs an extra smoothing operation over time, so that the
values of $\cL(h)(f)$ at times $t,t'$ differ by atmost $|t-t'|$.
We can show that if $J(m)(f, g) < \epsilon$, then $\forall
h.\forall t. |\cL(h)(f)(t) - \cL(h)(g)(t)| < 2\epsilon$.   We will
use this to establish non-convergence of function sequences in the
$J$-metric in the next two examples.

The first non-example shows that jumps are detected --- a sequence
of functions with jumps cannot converge to a continuous function.
\begin{example}\label{oneJNot}
Let $\{ b'_n \}$ be an increasing sequence that converges to
$\frac{1}{2}$.  Consider:

\[ f_{b'_n}(r) = \left\{
                \begin{array}{ll}
                0, &r < b'_n \\
                1, &b'_n \leq r < \frac{1}{2} \\
                0, &\frac{1}{2} \leq r
                \end{array}
                \right.
\]
These are depicted in the picture below.

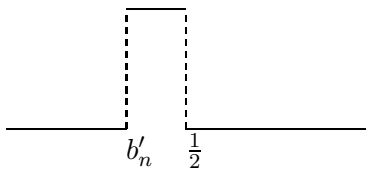
\setlength{\unitlength}{0.00625in}
\begin{picture}(300,200)(0,30)
\put(0,100){\line(1,0){100}}
\multiput(100,100)(0,10){10}{\line(0,1){5}}
\multiput(150,100)(0,10){10}{\line(0,1){5}}
\put(100,200){\line(1,0){50}}
 \put(150,100){\line(1,0){150}}
 \put(150,75){$\frac{1}{2}$}
 \put(100,75){$b'_n$}
\end{picture}

The sequence $\{ f_{b'_n} \}$ does not converge to the constant
$0$ function ${\mathbf 0}$, as $J(m)(f_{b'_n},{\mathbf 0}) = 1$,
because the point $(b'_n, 1)$ has distance 1 from the graph of
${\mathbf 0}$.  Alternatively, this can be illustrated by considering
the Lipschitz operator $h'_{\epsilon}$ on $[0,1]$ defined as:
\[ h'_{\epsilon}(r) = \max(0, \epsilon - |1 - r|) \]
For all $n$, $\cL(h'_{\epsilon})(g_n)(\frac{1}{2})  = \epsilon$,
but $\cL(h'_{\epsilon})({\mathbf 0})(\frac{1}{2}) = 0$.

\end{example}

The second non-example shows that continuous functions do not
approximate a function with jumps.
\begin{example}\label{contNot}~\cite{Whitt02}
Let $\{ \epsilon_n \}$ be an decreasing sequences that converges
to $0$.  Let $d_n = \frac{1}{2} - \frac{\epsilon_n}{2}$, $e_n\frac{1}{2} + \frac{\epsilon_n}{2}$.  Consider:

\[ h_{n}(r) = \left\{
                \begin{array}{ll}
                0, &r < d_n \\
                \frac{r - d_n}{\epsilon_n}, &d_n \leq r < e_n  \\
                1, &r \geq e_n
                \end{array}
                \right.
\]
These are depicted in the picture below.

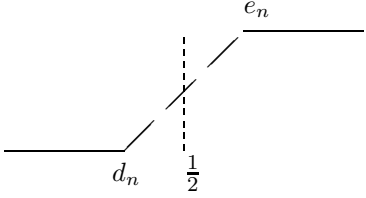
\setlength{\unitlength}{0.00625in}
\begin{picture}(300,200)(0,30)
\put(0,100){\line(1,0){100}}
 \put(200,200){\line(1,0){100}}
 \multiput(100,100)(35,35){3}{\line(1,1){25}}
\multiput(150,100)(0,10){10}{\line(0,1){5}}
 \put(150,75){$\frac{1}{2}$}
 \put(90,75){{\small $d_n$}}
 \put(200,215){{\small $e_n$}}
\end{picture}

The sequence $\{ g_{n} \}$ does not converge to $f_{\frac{1}{2}}$.
The point $(1/2, 1/2)$ is at distance $1/2$ from the graph of
$f_{\frac{1}{2}}$, so $J(m)(h_n, f_{\frac{1}{2}}) \geq 1/2$.
To analyze in terms of the operator $\cL$, consider the Lipschitz
operator $h_{\epsilon}$ on $[0,1]$ defined as:
\[ h_{\epsilon}(r) = \max(0, \epsilon - |\frac{1}{2} - r|) \]
For all $n$, $\cL(h_{\epsilon})(g_n)(\frac{1}{2}) = \epsilon$, but
$\cL(h_{\epsilon})(f_{\frac{1}{2}})(\frac{1}{2}) = 0$.
\end{example}

We conclude this section with a discussion of a delay operator on
the space of cadlag functions.
\begin{definition}
Let $(M,m)$ be a metric space.  Let $f \in \cadlag{(M,m)}$.  Let
$s \in M, 0 \leq r$.   Let $u:[0,r) \rightarrow M$ be a continuous
function.  Define $\delay{u}(f) \in \cadlag{(M,m)}$ as follows:
\[ \delay{u}(f)(t) = \left\{
                            \begin{array}{ll}
                            f(t-r), &\ t \geq r \\
                            u(t), &0 \leq t < r
                            \end{array}
                         \right.
\]
\end{definition}

The distance between a cadlag function and its $t$-delayed version
is no greater than $t$.
\begin{lemma}\label{shift}
Let $0 \leq r$.  Let $u:[0,r) \rightarrow M$ be continuous such
that  $(\forall 0 \leq r' < r) \ m(u(r'),f(0)) \leq r$. Then:
$J(m)(\delay{u}(f),f) \leq r $.
\end{lemma}\fullpaper{
\proof{} Follows from $m(\delay{u}(f)(t), f(t-r)) \leq r$, forall
$t$. \qed}

\section{GSMPs and cadlag functions.}\label{sec-cadl}

We deal with the temporal aspects of GSMPs next by constructing
cadlag functions for paths: i.e.\ sequences of generalized states
of a GSMP.


A sequence of generalized states is \emph{finitely varying} if it
is non-Zeno, i.e. for any i, $ \sum^{\infty}_{j>i}
\timeleft(\genstate{s_j}{c_j})$, the sum of the times spent at
each generalized state, diverges. Any finitely varying sequence of
generalized states $\genstate{s_i}{c_i}$ generates $f: [0, \infty)
\rightarrow \cG$ as follows:
\begin{itemize}
\item $f(t) = \genstate{s_i}{c}$, where
\[ \sum_{k=0}^{i} \timeleft(\genstate{s_k}{c_k}) \leq t <  \sum_{k=0}^{i+1} \timeleft(\genstate{s_k}{c_k}) \]
and $\clock{c} = \clock{c_i} - \rate{r_c}|t - \sum_{k=0}^{i}
\timeleft(\genstate{s_k}{c_k}) |$ is the new clock values after
evolving at rate $\rate{r_c}$ for time $|t - \sum_{k=0}^{i}
\timeleft(\genstate{s_k}{c_k}) |$ starting from
$\genstate{s_k}{c_k}$.
\end{itemize}
Such finitely varying traces satisfy the following: for any
interval $[t,t']$, that there is a finite partition $t=t_0 < t_1 <
t_2 \ldots <t_n =t'$ such that:
\begin{itemize}
\item If $f(t_i) = \genstate{s}{c}$, then: $(\forall t_i \leq t < t_{i+1})
f(t) = \langle s, \clock{c} - \rate{r_c} |t-t_i| \rangle$.
\end{itemize}
We write $\traces{\genstate{s}{c}}$ for the set of traces that
start with $\genstate{s}{c}$.  The probability distributions
associated with initial clock-values at states ($f_s$) and
transitions ($\next$) induces a probability measure on
$\traces{\genstate{s}{c}}$.  The paths that are Zeno have measure
zero, so the finitely-varying paths generate
$\traces{\genstate{s}{c}}$ in measure.

Arbitrarily close approximations to the distance between
finitely-varying functions $f,g \in \cadlag{(\cG,m)}$ are forced
by the distances between the values of $f,g$ at finitely many
points of time.  This lemma is useful later on to show that our
coinductive definition of metric has closure ordinal $\omega$.
\begin{lemma}\label{random2}
Let $m$ be a pseudometric over generalized states.   Let $f,g \in
\cadlag{(\cG,m)}$ be finitely-varying functions such that
$J(m)(f,g) > \delta$.

Then there is a finite subset $\cG_{fin} \subseteq \cG$ and
$\epsilon
>0$ such that for any $m'$, $J(m')(f,g) > \delta$ if
$(\forall \genst{1},\genst{2} \in \cG_{fin}) \
[m'(\genst{1},\genst{2}) \geq m(\genst{1},\genst{2}) - \epsilon]$.
\end{lemma}
\proof{} If $J(m)(f,g) > \delta$, without loss of generality there
is a $t$ such that the Hausdorff distance of $(f(t),t)$ from
$\Gr{g}$ is greater than $\delta + \gamma$ for some $\gamma >0$.
Consider the bounded interval $[t-\delta, t+ \delta]$.  By
finite-variation of $g$, we have a partition $t_0 = t-\delta < t_1
< t_2 \ldots <t_n =t + \delta$ such that:
\begin{itemize}
\item $(\forall t_i \leq t < t_{i+1})
f(t) = \langle s, \clock{c} - \rate{r_c} (t-t_i) \rangle $, where
$f(t_i)=\genstate{s}{c}$
\item $|t_i - t_{i+1}| < \frac{\gamma}{2}$
\item If $t',t''$ are in the same
partition, then \\
$m(f(t'), f(t'')) < \frac{\gamma}{2}$
\end{itemize}
Construct $\cG_{fin} = \{ \genstate{s_1}{c_1}, \ldots,
\genstate{s_n}{c_n} \}$ by choosing one $\genstate{s_i}{c_i}$ each
from $g[t_{i-1},t_i]$ Choose $\epsilon = \frac{\gamma}{2}$.  For
any $m'$ on generalized states such that:
\[ (\forall i,j) m'(\genstate{s_i}{c_i},\genstate{s_j}{c_j}) \geq
m(\genstate{s_i}{c_i},\genstate{s_j}{c_j}) - \epsilon \] the
Hausdorff distance of $(f(t),t)$ from $\Gr{g}$ is greater than
$\delta $ by construction. \qed

\section{Bisimulation style definition of metric}\label{sec5}

Let $\met$ be the class of pseudo-metrics on generalized states
that satisfy:
\[ \mathtt{Props}(s) \not = \mathtt{Props}(s') \Rightarrow
m(\genstate{s}{c},\genstate{s'}{c'}) = 1 \] where
$\mathtt{Props}(s)$ is the set of atomic propositions true in
state $s$.  We order these pseudo-metrics as in
section~\ref{sec2}: $m_1 \preceq m_2 \ \mbox{ if } (\forall s,t)
m_1(s,t) \geq m_2(s,t) $.  Fix $0 < k <1$.  Define a functional
$\cF$ on $\met$:
\begin{definition}
$\cF(m)(\genstate{s}{c},\genstate{s'}{c'}) < \epsilon$ if
\[ k \times W(J(m))(\traces{\genstate{s}{c}},
\traces{\genstate{s'}{c'}}) < \epsilon
\]
\end{definition}
In this definition, view $k$ as a discount factor.  In the next
section, we will show that the choice of $k$ does not affect the
essential character of the metric.  ``Type-checking'' of this
definition provides some intuitions: $m$ is a pseudo-metric on
generalized states.  $J(m)$, following Skorohod J2,  is a
pseudo-metric on finitely varying sequence of generalized states.
$W(J(m))$, following Wasserstein,  is a pseudo-metric on
probability distributions on finitely varying sequence of
generalized states.

As an immediate consequence of lemmas~\ref{sk-mon}
and~\ref{wass-mon}:
\begin{lemma}
$\cF$ is monotone on $\met$.
\end{lemma}
Since $( \met,\preceq)$  is a complete lattice, $\cF$ has a
maximum fixed point, $m_{\cF}$.
\begin{definition}
$m$ is a metric-bisimulation if $m  \preceq \cF(m)$.
\end{definition}
It is well known that the greatest fixed point of $\cF$ is given
by:
\[ m_{\cF} = \bigsqcup \{ m \mid  m \mbox{ is a metric-bisimulation} \}  \]
Thus, a metric-bisimulation $m$ provides an upper bound on the
distances assigned by $m_{\cF}$.

Consider the equivalence relation $\simeq  = \{ (\genstate{s}{c},
\genstate{s'}{c'}) \mid m_{\cF}(\genstate{s}{c},
\genstate{s'}{c'}) = 0 \}$.  $\simeq$  describes a notion of
bisimulation and is explicitly defined as follows.

Let $\met_{ \{0,1 \}}$, the sublattice of $\met$ consisting of
metrics whose range is $\{0,1\}$, i.e.\ all distances are either
$0$ or $1$. $\met_{ \{0,1\}}$ is essentially the class of
equivalence relations.  A simple proof shows that:

\[ \simeq = \bigcup \{ m^{-1}(0) \mid  m \in \met_{ \{0,1 \}}, m  \preceq
\cF(m) \} \]

As an example of metric-reasoning, we now show that generalized
states with the same state, but clock values reflecting evolution
for a time $t$ are $m_{\cF}$-close.

\begin{lemma}\label{basic}
Define a pseudo-metric $m$ on generalized states as follows:
\[
m(\genstate{s}{c},\genstate{s'}{c'}) = \left\{
  \begin{array}{ll}
       \min(1,t), &\mbox{ if } s = s' \ \mbox{ and } \\
                  &\clock{c} = \clock{c'} + \rate{r_c} t \  \mbox{or} \\
                  &\clock{c} = \clock{c'} - \rate{r_c} t     \\
       1, &\mbox{otherwise}
  \end{array} \right.
\]
Then: $m \preceq \cF(m)$.
\end{lemma}

\proof{} It suffices to prove that for all generalized states
$\genstate{s}{c},\genstate{s'}{c'}$
\[
\cF(m)(\genstate{s}{c},\genstate{s'}{c'}) \leq
m(\genstate{s}{c},\genstate{s'}{c'}). \]
The only case to consider
is when $s=s'$, wlog assume $\clock{c} \leq \clock{c'}$.

Define $u:[0, t) \rightarrow \cG$ as follows.
\[ u(t) = \langle s, \clock{c'} - \rate{r_c} t \rangle \]
With this definition, it is clear that:
\[ \traces{\genstate{s'}{c'}} = \{ \delay{u}(f) \mid f \in
\traces{\genstate{s}{c}} \} \] with the distribution inherited
from $\traces{\genstate{s}{c}}$. Now this induces a matching
between the traces in $\traces{\genstate{s}{c}}$ and those of
$\traces{\genstate{s'}{c'}}$.  This in turn induces a distribution
on the product space $\traces{\genstate{s}{c}}\times
\traces{\genstate{s'}{c'}}$.  Using this distribution as the
$\rho$ in the dual form of the definition  we get the result. \qed

Let $m_0 = \top$ and $m_{i+1} = \cF(m_i)$.  The role played by the
discount constant $k$ is captured in the following fact:
\[ (\forall \genstate{s}{c}, \genstate{s'}{c'}) \
|m_{n+1}(\genstate{s}{c}, \genstate{s'}{c'}) -
m_n(\genstate{s}{c}, \genstate{s'}{c'})| \leq k^{n+1}. \]
This is
the key step in the proof of the following lemma.

\begin{lemma}\label{mf_is_sep}
$m_{\cF}$ is separable.
\end{lemma}
\proof{} We first note that if $m,m'$ are such that  $(\forall s)
|m(s)-m'(s)| \leq \delta$, then:
\begin{itemize}
\item $(\forall P,Q) \ |W(m)(P,Q) - W(m')(P,Q)| \leq \delta$, this is
immediate from the dual formulation
\item  $(\forall f,g) |J(m)(f,g) - J(m')(f,g)| \leq \delta$,
this is immediate from the definition.
\end{itemize}

We prove by induction on $n$ that 
\[(\forall \genstate{s}{c},
\genstate{s'}{c'}) \ |m_{n+1}(\genstate{s}{c}, \genstate{s'}{c'})
- m_n(\genstate{s}{c}, \genstate{s'}{c'})| \leq k^{n+1}.\]

\begin{itemize}
\item Base. $n=0$.  Follows from the fact that $m_0$ is the constant $0$
function and $m_1$ is bounded above by $k$.

\item Induction.  Assume $(\forall \genstate{s}{c}, \genstate{s'}{c'})$
$|m_{n+1}(\genstate{s}{c}, \genstate{s'}{c'}) -
m_n(\genstate{s}{c}, \genstate{s'}{c'})| \leq k^{n+1} $. Thus, for
any $\genstate{s}{c}, \genstate{s'}{c'})$, $W(J(m_{n+1})$ and $
(\genstate{s}{c}, \genstate{s'}{c'})$ and $W(J(m_n))$ differ by
atmost $k^{n+1}$. So:\[ |m_{n+2}(\genstate{s}{c},
\genstate{s'}{c'}) - m_{n+1}((\genstate{s}{c}, \genstate{s'}{c'})|
\leq k^{n+2} \]

\end{itemize}
Let $m = \sup m_i$.  From above, $(\forall \genstate{s}{c},
\genstate{s'}{c'})$
\[ |m(\genstate{s}{c}, \genstate{s'}{c'}) -
m_n(\genstate{s}{c}, \genstate{s'}{c'})| \leq \frac{k^{n+1}}{1-k}
\]
Thus, an $\epsilon$ ball around $\genstate{s}{c}$ wrt the metric
$m$ can be realized as the countable union of open sets wrt the
metrics $m_n$.  The result now follows from the separability of
the metrics $m_n$.

\qed

The separability of $m_{\cF}$ enables one to prove the analogue of
lemma~\ref{random2}.
\begin{lemma} [Finite detectability of distances]\label{finite} \hfill \\
Let $m$ be a pseudometric on $\cG$ with countable basis. Let
$\cF(m)(\genstate{s}{c},\genstate{s'}{c'}) > \delta$.

Then there is a finite subset $\cG_{fin} \subseteq \cG$ and
$\epsilon > 0$ such that for any metric $m' \succeq m$,
$\cF(m')(\genstate{s}{c},\genstate{s'}{c'}) >
\delta$ if $(\forall \gs,\gs' \in \cG_{fin}) \ [m'(\gs,\gs') \geq
m(\gs,\gs') - \epsilon]$.
\end{lemma}
\fullpaper{
\proof \hfill \\
Let $\cF(m)(\genstate{s}{c},\genstate{s'}{c'}) > \delta +
\gamma$, for $\gamma > 0$.

From separability of $m$, lemma~\ref{cadlag-sep} yields
separability of $J(m)$.  Let $\rho$ be the measure induced on the
space of all traces by
$\traces(\genstate{s}{c})$. Using separability of $J(m)$, we can get a
finite partition $U_0, U_1,\dots, U_n$ of $\cG$ satisfying ${\tt diameter}
(U_i) < \frac{\gamma}{16}$ for $i \geq 1$, and $\rho(U_0) <
\frac{\gamma}{16}$.   Using lemma~\ref{random1} with $\epsilon \frac{\gamma}{16}$ gives
us a finite set of traces $L_1 = \{f_i\mid f_i \in U_i, i \geq 0\}$ with
probabilities $p_i = P(f_i) = \rho(U_i)$.  Similarly
applying lemma~\ref{random1} to  $\traces(\genstate{s'}{c'})$ gives us
another finite
set $L_2 = \{f'_i\mid i\}$ with
probabilities $p'_i$ given by the measure induced by
$\traces(\genstate{s'}{c'})$.  We then have from lemma~\ref{random1}

\begin{itemize}
\item $W(J(m)) (\traces(\genstate{s}{c}),L_1) < \frac{\gamma}{8}$
\item $W(J(m)) (\traces(\genstate{s'}{c'}),L_2) < \frac{\gamma}{8}$
\end{itemize}
So, $W(J(m))(L_1,L_2) > \delta + \frac{3\gamma}{4}$.

Using corollary~\ref{random2} for every pair in $\{ f_i \mid i\}
\times \{ f'_j \mid j \}$, yields a
finite set $\cG_{fin} \subseteq \cG$ and $\epsilon
>0$ such that for any $m'$, if $(\forall \gs,\gs' \in \cG_{fin}) \
[m'(\gs,\gs') \geq m(\gs,\gs') - \epsilon]$, then $J(m')(f_i,f'_j)
> J(m)(f_i,f'_j) - \frac{\gamma}{4}$ for all $i,j$.
A simple use of the dual form of the Wasserstein metric
yields $W(J(m'))(L_1,L_2) > W(J(m))(L_1,L_2) - \frac{\gamma}{4}$
under these conditions.

Now since $m' \succeq m$, $W(J(m')) (\traces(\genstate{s}{c}),L_1)
\leq W(J(m)) (\traces(\genstate{s}{c}),L_1) < \frac{\gamma}{8}$,
and similarly $W(J(m')) (\traces(\genstate{s'}{c'}),L_2) <
\frac{\gamma}{8}$. The triangle inequality then gives us that
$\cF(m')(\genstate{s}{c},\genstate{s'}{c'}) > \delta$. \qed}

\begin{lemma}
$\cF$  has closure ordinal $\omega$.
\end{lemma}
\proof{} The proof proceeds by showing that the maximum fixed
point $m$ is given by $m = \sqcup_i m_i$, where $m_0 = \top$ and
$m_{i+1} \cF(m_i)$.

Let $m(\genstate{s}{c}, \genstate{s'}{c'}) > \delta$.  From
lemma~\ref{finite}, we deduce the finitely many conditions of the
form
\[ m'(\genstate{s_i}{c_i}, \genstate{s'_i}{c'_i}) > m(\genstate{s}{c},
\genstate{s'}{c'}) - \epsilon \] that suffice to ensure that
$\cF(m')(\genstate{s}{c}, \genstate{s'}{c'}) > \delta$.  Each of
these finitely many conditions are met at a finite index,
therefore by $\omega$ they are all met and the result follows.
\qed

\section{Uniform spaces}\label{sec8}

A metric captures a quantitative notion of distance or
``nearness'', a topology captures a qualitiative notion of
nearness: a topology is enough to talk about convergence and
continuity.  A topology is, however, not enough to capture a
notion of relative distance.  One cannot say ``x is closer to $y$
than it is to $z$'' on the basis of a topology alone.  A uniform
space -- see, for example,~\cite{Geroch85} for a quick survey --
captures the essence of the relative distance notion in metric
spaces: if there are points $x,y,z$ such that $x$ is closer to $y$
than to $z$, uniform spaces have enough data to capture this
without committing to the actual numerical values of the
distances.  The aim of this section is to show that our treatment
is ``upto uniformity'' -- this is a formal way of showing that
there is no ad-hoc treatment of the quantitative metric distances.
In particular, we show that different discount factors $k$ yield
the same uniformity.

Let $S$ be a set.
\begin{definition}
A \emph{pseudo-uniformity}, $\cU$ is a collection of subsets of $S
\times S$, called entourages, that satisfies:
\begin{itemize}
\item $(\forall E\in \cU) \ (\forall x \in S) \ (x,x) \in E$
\item $E \in \cU \Rightarrow\ E^{-1} \in \cU$
\item $E \in \cU \Rightarrow\ (\exists E' \in \cU) \ E' E'
\subseteq E $
\item $E, E' \in \cU \Rightarrow\ E \cap E' \in \cU$
\item $E \in \cU, E \subseteq E'  \Rightarrow\ E' \in \cU$
\end{itemize}
\end{definition}
One can think of the entourages as defining approximations to the
identity relation - just as the neighbourhood of a point can be
thought of as an approximation of the point.  The first axiom says
this, the second axiom is symmetry and the third is a truncated
version of transitivity and the final axiom says that a superset
of an approximation to the identity is also an approximation to
the identity.  The usual presentation of uniformities also
includes $\bigcap_{E \in \cU} E = \{ (x,x) \mid x \in S \}$, but
this condition is not appropriate to our pseudo-metric setting. In
this paper, we will work with pseudo-uniformities, often dropping
``pseudo'' and merely saying ``uniformities''.

\begin{definition}
A pseudo-uniform space is a pair $(\cS,\cU)$ where $\cU$ is a
pseudouniformity on $\cS$.
\end{definition}

There is a natural notion of map between uniform spaces.  A
morphism between uniform spaces generalizes uniformly-continuous
functions.
\begin{definition}
A morphism $f$ between (pseudo)uniform spaces $f: (S_1,\cU_1)
\rightarrow  (S_2,\cU_2)$ is a function $f: S_1 \rightarrow S_2$
such that
\[ (\forall E_2 \in \cU_2) \hat{f}^{-1}(E_2) \in \cU_1 \]
where $\hat{f}:S_1\times S_1\to S_2\times S_2$ is given by
$\hat{f}(x,y)=(f(x),f(y))$.
\end{definition}

To gain intuition into this definition, we describe how a
pseudometric generates a pseudo-uniformity.  Given a pseudo-metric
$m$ on $\cS$, let $K^{\epsilon}_{m} = \{ (x,y) \mid m(x,y) <
\epsilon \}$ for $\epsilon > 0$. We get a pseudo-uniformity by
considering~\cite[p.218]{Geroch85}:
\[ \cU = \{ E \subseteq \cS \times \cS \mid (\exists \epsilon)
K^{\epsilon}_{m} \subseteq E \} \] Thus, if $m(x,y) < \epsilon$
and $m(x,z) >\epsilon$, there is an entourage that contains
$(x,y)$ but not $(x,z)$.  Clearly if one just scales the metric this
construction yields the same uniformity.

Thus, all pseudometrics induce pseudo-uniformities, but the
converse is not true: there are pseudo-uniformities that are not
induced by metrics.  Two metrics $m,m'$ on the same set induce the
same uniformity if and only if the identity map is uniformly
continuous in both directions.

\subsection{The lattice of uniformities}
Consider uniformities induced by pseudo-metrics on a fixed set of
states $\cS$.
\begin{definition}
$\pmu$ is the class of pseudo-metrizable uniformities $\{ \cU_i
\}$ on $\cS$ ordered as follows.
$$\cU_1 \leq \cU_2 \ \mbox{ if } \  \cU_2 \subseteq \cU_1$$
\end{definition}
This order is closely related to the order on the lattice of
pseudometrics.
\begin{lemma}
Let pseudometrics $m_1,m_2$ induce $\cU_1,\cU_2$ respectively, Let
$\cU_2 \subseteq \cU_1$.  Then pseudometric $m$ defined as $m(s,t)
= \max(m_2(s,t),m_1(s,t))$ also induces $\cU_1$.
\end{lemma}
\proof{} Let $\cU$ be the uniformity induced by $m$.  We need to
show that $\cU = \cU_1$.  Since $K^{\epsilon}_{m} \subseteq
K^{\epsilon}_{m_1}$, $\cU \supseteq \cU_1$.

We now show that $\cU \subseteq \cU_1$.  Consider an entourage $E
\in \cU$.  Thus, there is an $\epsilon$ such that $E \supseteq
K^{\epsilon}_{m}$, i.e. $E \supseteq K^{\epsilon}_{m_1} \cap
K^{\epsilon}_{m_2}$.  But $K^{\epsilon}_{m_2} \in \cU_2$ and by
assumption $\cU_2 \subseteq \cU_1$, $K^{\epsilon}_{m_2} \in
\cU_1$.  So, $K^{\epsilon}_{m_1} \cap K^{\epsilon}_{m_2} \in
\cU_1$.  So, $E \in \cU_1$. \qed

\begin{lemma}
$( \pmu,\leq)$  is a complete lattice.
\end{lemma}
\proof{} The least element is given by the discrete metric: $
\bot(s,t) = 0 \mbox{ if } s = t, 1 $ otherwise. The top element
has only one entourage $\cS$ and is induced by the pseudometric
given by $(\forall s,t) \top(s,t) =0$. The greatest lower bounds
of $\{ \cU_i \}$ is given by the $\cup_i \cU_i$. \qed

\subsection{Wasserstein uniformity}
\begin{lemma}\label{w-uniform}
Let $(M,m), (M,m')$ be such that the uniformities induced by
$m,m'$ are the same.  Then, $W(m), W(m')$ induce the same
uniformity.
\end{lemma}

\proof{} Since the uniformities induced by $m,m'$ are the same,
the identity map $(M,m) \rightarrow(M,m')$ is uniformly
continuous. We need to show that the identity map on distributions
with metrics $W(m)$ and $W(m')$ is uniformly continuous.

Since the uniformity induced by $W(m)$ (resp. $W(m')$) is the same
as the uniformity induced by the Prohorov metric $\pi(m)$ (resp.
$\pi(m')$), it suffices to prove that the identity map on
distributions with metrics $\pi(m)$ and $\pi(m')$ is uniformly
continuous.

Let $\pi(m')(P,Q) \leq \epsilon$.  Let $\delta_{\epsilon}$ be such
that: $m(x,y) < \delta_{\epsilon} \Rightarrow m'(x,y) < \epsilon$.
Let $\gamma = \min(\epsilon, \delta_{\epsilon})$.  Then:
\begin{eqnarray*}
\pi(m)(P,Q) \leq \gamma &\Rightarrow& P(U) \leq Q(U^{\gamma}_m) +
\gamma \\
&\Rightarrow& P(U) \leq Q(U^{\epsilon}_{m'}) +
\gamma \ \  [Q(U^{\gamma}_m) \subseteq Q(U^{\epsilon}_{m'})] \\
&\Rightarrow& P(U) \leq Q(U^{\epsilon}_{m'}) + \epsilon \ \ (
\gamma \leq
\epsilon) \\
&\Rightarrow& \pi(m')(P,Q) \leq \epsilon
\end{eqnarray*}
\qed

In the light of this theorem, we write $W(\cU)$ for the uniformity
generated by a pseudo-metrizable uniformity $\cU$. As a direct
consequence of lemma~\ref{wass-mon}, we have:
\begin{corollary}
$\cU \leq \cU' \Rightarrow W(\cU) \leq W(\cU')$.
\end{corollary}

\subsection{Skorohod J2 uniformity}
A similar result holds for the Skorohod metric.
\begin{lemma}\label{sk-uniform}
The uniformity induced by the Skorohod J2 metric depends only the
uniformity induced by $m$ on $M$.
\end{lemma}
\proof{} For each $ f \in \cadlag{M}$ consider its graph, $ \Gr{f}
= \{ (f(t),t) \}$.  For a relation $R$ and a set $X$, write $R(Y)
= \{ x \mid (x,y) \in R, y \in  Y \}$.  Similarly, $ (X) R = \{ y
\mid (x,y) \in R, x \in  X \}$.  These operations are monotone,
under the subset ordering, in both $X$ and $R$.

Let the space $\cS = [0,\infty) \times M$ be equipped with the
metric $d((x,s), (y,t)) = max(|x-y|, m(s,t))$.  This metric
induces a uniformity $U(\cS)$ on $\cS$.

Let $E \subseteq \cS \times \cS$ be an entourage in $U(\cS)$.
Consider the subset of $J(E)$ of $\cadlag{M} \times \cadlag{M}$
induced by $E$ as follows.  $J(E)$ is the set of all $(f,g)$ such
that:
\begin{itemize}
\item  $ (\Gr{f}) E  \supseteq \Gr{g}$
\item  $ E (\Gr{g}) \supseteq \Gr{f}$
\end{itemize}
Consider $\cU = \{ S \mid S \supseteq J(E), E \subseteq \cS \times
\cS, E \in U(\cS) \}$. We will show that $\cU$ is the uniformity
generated by $J(m)$.

Rewriting $J(m)$ in this style: $J(m)(f,g) < \epsilon
\Leftrightarrow$
\begin{itemize}
\item  $ (\Gr{f}) K^{\epsilon}_{\langle m,
|\cdot| \rangle}  \supseteq \Gr{g}$
\item  $ K^{\epsilon}_{\langle m,
|\cdot| \rangle} (\Gr{g}) \supseteq \Gr{f}$
\end{itemize}
Clearly, the uniformity generated by $J(m)$ is a subset of $\cU$,
since $K^{\epsilon}_{\langle m, |\cdot| \rangle}$ is an entourage
of $\cS \times \cS$, and thus considered in the definition of
$\cU$.  Furthermore, any arbitrary entourage $E$ in $U(\cS)$ is a
superset of these basic entourages, and yields a superset by
monotonicity of the relational operations.

The result now follows by the upward-closure axiom in the
definition of $\cU$ and the uniformity generated by $J(m)$. \qed
The proof of this theorem also shows that the construction $J(m)$
yields the same uniformity for other definitions of metrics on $M$
that yield the same uniformity, e.g. the metric $d((x,s), (y,t))
|x-y| + m(s,t))$.

In light of this theorem, we write $J(\cU)$ for the uniformity
generated by a pseudo-metrizable uniformity $\cU$.  As a direct
consequence of lemma~\ref{sk-mon}, we have:
\begin{corollary}
$\cS_1 \leq \cS_2 \Rightarrow J(\cS_1) \leq J(\cS_2)$
\end{corollary}

\subsection{A functional on the lattice of uniformities}
Combining the above results, we deduce the existence of a monotone
function $\cF$ on the lattice of uniformities.  This function is
insensitive to the actual numerical value of the discount factor
$k$.
\begin{lemma}
$(\forall 0 < k,k' \leq 1) \ {\mathcal F}_k = {\mathcal F}_{k'}$.
\end{lemma}
\proof{} By Lemmas~\ref{w-uniform} and~\ref{sk-uniform} we see
that the functional $\cF$ is defined upto uniformity.  If we
change $k$ to $k'$ we are simply rescaling the metric, this
clearly gives the same uniformity. \qed

Furthermore, for any discount factor $0< k <1$, we get the same
maximum fixed point in the lattice of uniformities.  In contrast
to the above lemma, the following theorem relies on $k \not=1$.
\begin{theorem}\label{maximal-uniform}
The maximum fixpoint in $(\pmu,\leq)$ is the uniformity induced by
$m_{\cF}$, the maximum fixpoint in $(\met,\preceq)$.
\end{theorem}
\proof{} It suffices to show that the greatest lower bound of the
$\cU_i$, $\cup \cU_i$ is the uniformity induced by $\sup m_i$.
Let $\mathcal{V}$ be the uniformity induced by $\sup m_i$.  Since
$m_i \preceq \sum m_i$ for all $i$, we have that $\cU_i \subseteq
\mathcal{V}$ for all $i$.  Thus $\cup \cU_i \subseteq \mathcal{V}$
and hence the identity function from $\mathcal{V}$ to $\cU_i$ is
uniformly continuous.

For the converse, we will show that the identity function from
$\cup \cU_i \to \mathcal{V}$ is uniformly continuous.  We know -
from the proof of Lemma~\ref{mf_is_sep} - that for any generalized
states $\gs$ and $\gs'$
\[ \sup m_i(\gs,\gs')  - m_i(\gs,\gs')  \leq \frac{k^{n+1}}{1-k}. \]
We choose $n$ and $\delta$ such that $\delta + \frac{k^{n+1}}{1-k}
< \epsilon$.  Now for any such $\delta$ and $n$ we have
\[  m_n(\gs,\gs') < \delta \Rightarrow \sup m_i (\gs,\gs') < \epsilon.\]
This shows that for any $K^{\epsilon}_{\sup m_i}$ is contained in
a $K^{\delta{}}_{m_n}$ and hence that the identity function is
uniformly continuous. \qed This proof relies on the fact that
$k<1$ otherwise the $\delta{}$ would not be defined.

\section{Examples}\label{sec7}

In this section, we discuss several examples of the use of
approximate reasoning techniques.  The general approach in this
section is to identify natural quantitative observables, already
explored in the literature, that are amenable to approximation ---
i.e. to calculate the observable at a state $\genstate{s}{c}$ upto
$\epsilon$, it suffices to calculate it a close-enough state
$\genstate{s'}{c'}$.  This is clearly implied by continuity of the
observable w.r.t. the metric $m_{\cF}$.

The main technical tool that we use to establish continuity of
observables is a continuous mapping theorem, e.g.
see~\cite{Whitt02,Shedler87} for an introductory exposition.
\begin{theorem}[Continuous mapping theorem] \hfill \\
Let $P_n$ be a sequence of probability distributions on $X$ that
weakly converge to $P$.  Let $U$ be a continuous function $X \rightarrow
R$.  Then $\int U dP_n$ converges to $\int U dP$.
\end{theorem}

\subsection{Expected time to hit a proposition}

Let $\mathbf{p}$ be a proposition.  We consider the expected time
required to hit a $p$-state, i.e. a state where the proposition
$\mathbf{p}$ is true. Define ${\tt Hit}_p: \cadlag{m_F}
\rightarrow [0,\infty]$:
\[ {\tt Hit}_p(f) = \inf \{ t \mid f(t) = \genstate{s}{c}, \mathbf{p} \
\mbox{ true at } s \} \]

${\tt Hit}_p$ is a continuous function --- if $J(m_F)(f,g) <
\epsilon$, then $|{\tt Hit}_p(f) - {\tt Hit}_p(g)| < \epsilon$.

So, using the continuous mapping theorem, we deduce that if
$\{\genstate{s_i}{c_i} \}$ converges to $\genstate{s}{c}$ then the
sequence of expected times to hit a $p$-state from
$\{\genstate{s_i}{c_i} \}$ converges to the expected time to hit a
$p$-state from $\{\genstate{s}{c} \}$.  In fact, since in this
case, ${\tt Hit}_p$ is a $1$-Lipschitz function, we can also
deduce the rate of convergence using~\cite{Whitt02} --- if
$m_F(\genstate{s_i}{c_i}, \genstate{s}{c}) < \epsilon$, then the
expected times to hit a $p$-state differ by atmost $2 \epsilon$.

\subsection{Expected rewards}
Let $r_i$ be an assignment of rewards to states $s_i$ such that if
$r_i \not =r_j$, then states $s_i$, $s_j$ differ in the
truth-assignment of at least one proposition.  This restriction
can be viewed purely as a modelling constraint.

Define a function $R: \cG \rightarrow [0, \infty)$ by:
\[ R(\genstate{s_i}{c} = r_i \]
Under the hypothesis that distinct rewards are distinguished
propositionally, $R$ defines a continuous function.

For any finitely-varying $f$, consider $ {\tt CumR}(f)$, a
continuous function of $t$ defined as follows:
\[ {\tt CumR}(f)(T) = \int_0^T R(f(t)) dt \]

By standard results --- e.g. see~\cite{Whitt02} --- ${\tt CumR}$
is a continuous function from $\cadlag{m_F}$ to $(C,{\tt unif})$
where $C$ is the space of continuous functions from $[0, \infty)
\rightarrow [0, \infty)$ with the uniform metric:
\[ {\tt unif}(f,g) = \sup_t |f(t) - g(t)| \]

Consider the following continuous functions from $(C,{\tt unif})$
to $[0, \infty)$:
\begin{itemize}
\item For a fixed $T$, cumulative reward at time $T$.
\item For a fixed $T$, average reward per unit time at $T$.
\item The supremum of the times $T$ at which cumulative reward is
less than a fixed $v$, for some value $v$.
\item  The supremum of the times $T$ at which the average reward is
less than a fixed $v$, for some value $v$.
\end{itemize}
In each of these cases, by composing with ${\tt CumR}$, we get a
continuous function from $\cadlag{m_F}$ to $[0,\infty)$.  So, the
continuous mapping theorem applies, and we deduce that if
$\{\genstate{s_i}{c_i} \}$ converges to $\genstate{s}{c}$ then the
sequence of expected values from $\{\genstate{s_i}{c_i} \}$
converges to the expected value at $\{\genstate{s}{c} \}$.

\section{Functional characterization of uniformity}\label{sec6}
In an early treatment of metrics for LMPs~\cite{Desharnais99b,Desharnais04}
the metric was defined through a class of functions closely related to a
modal logic.  The idea was that, in a probabilistic setting, random
variables play a role analogous to modal formulas.  A class of random
variables (measurable functions) was defined on the state space and the
metric was obtained by taking the sup over this class of functions.
The coinductive definition of the metric came
later~\cite{vanBreugel01a,Desharnais02a} and was shown to be the same as
the metric defined logically.  In the present work we develop the subject
along similar lines.  We already have the fixed-point version of the
metric: we now give the ``logical'' view.
In this section, we provide an explicit construction of the maximum fixed
point by considering a class of $[0,1]$ valued functions.

\subsection{Function expressions}
\begin{definition}\label{d:syntax}
\rm Fix $0 < k \leq \frac{1}{2}$.  The syntax of function
expressions is given by:
\begin{eqnarray*}
F_k &::=&  \mathbf{1}\mid \mathbf{p} \mid \min(F_k,F_k) \mid h\circ F_k\mid
\int G_k \\
G_k &::=&  \cL(F_k)(t) \mid \min(G_k(t),G_k(t')) \mid h\circ
G_k(t)
\end{eqnarray*}
where $\mathbf{p}$  ranges over atomic propositions, $h$ is any
Lipschitz function on $[0,1]$, $t \in [0, \infty)$.
\end{definition}
The subscript $k$ gives the
discount factor.  We will not usually write this factor explicitly.
Intuitively the F-function expressions are evaluated at
generalized states, and the G-function expressions are evaluated
on finitely-varying paths at the times shown.  In a temporal logic with
state and path formulas, like CTL*, the path formulas are \emph{implicitly}
evaluated at the first time.  This may not seem to be the case with a
formula like $Gp$ ($\Box p$ in LTL notation) but is clear with a formula
like $Xp$ ($\bigcirc p$).  In our $G$-formulas we cannot have a first time:
we provide the time as an explicit parameter.  One can imagine a much
richer language of path formulas: for example, one might have time averages
along a path.  However, the present language suffices for the definition of
the metric.

As preliminary intuition, $\mathbf{1}$ corresponds to the formula
{\tt true}, $\min(\cdot, \cdot)$ corresponds to
conjunction\footnote{$\max(\cdot, \cdot)$ is definable as $1 -
\min(1- \cdot, 1-\cdot)$ in both classes of function
expressions.}, and $h\circ f$ encompasses both testing (via $h(x)
= \max(x - q,0)$) and negation (via $h(x) = 1 - x$).  At a
generalized state $\genstate{s}{c}$, $\int G(t)$ yields the
(discounted) expectation of $G(t)$ wrt the distribution of
$\traces{\genstate{s}{c}}$.  The intuition underlying
$\cL(F)(t)$ has been discussed in section~\ref{sec3} --- at a
finitely-varying function $f$, $\cL(F)(t)$ yields the evaluation
at time $t$ of a time-smoothed variant of $f$.

We formalize these intuitions below.  The interpretations of
F-function expressions and G-function expressions yield maps
whose range is the interval $[0,1]$.
\begin{itemize}
\item The domain of F-function expressions is $\cG$, the set of
generalized states.

\item The domain of G-function expressions is the set
of finitely-varying functions with range $\cG$.
\end{itemize}
Fix a GSMP.  F-function expressions are evaluated as follows at a
generalized state $\genstate{s}{c}$:
\begin{eqnarray*}
\mathbf{p}(\genstate{s}{c}) &=& 1, \mbox{ iff } {\tt p} \ \mbox{true at } s \\
  \mathbf{1}(\genstate{s}{c}) &=& 1\\
\min(F_1,F_2)(\genstate{s}{c})&=&\min(F_1(\genstate{s}{c}),
F_2(\genstate{s}{c}))\\
        h\circ F(\genstate{s}{c})& =& h(F(\genstate{s}{c})) \\
( \int G(t) ) (\genstate{s}{c})& = & k \times \int G(t)(f) \mathrm{d}\mu
\end{eqnarray*}
where $\mu$ is the distribution of $\traces(\genstate{s}{c})$.  Note that
in this definition $f$ varies among the paths of $\traces(\genstate{s}{c})$
so $G(t)$ is a measurable function on the space of these paths and $\mu$ is
a measure on these paths.

G-function expressions are evaluated as follows at a
finitely-varying function $f$:
\begin{eqnarray*}
\cL(F)(t)(f) &=& \sup_{t'} \{ F(f(t')) - | t' -t| \} \\
\min(G_1(t_1),G_2(t_2))(f)&=&\min(G_1(t_1)(f),G_2(t_2)(f))\\
        h\circ G(t)(f)& =& h(G(t)(f))
\end{eqnarray*}
Thus, for $f \in \traces(\genstate{s}{c})$, $\cL(F)(t)(f)$ is
the upper Lipschitz approximation to $F_k^s \circ f$ evaluated at
$t$.
\subsection{A pseudometric from function expressions}
We define a pseudometric $d_k$ as follows.
\begin{definition} \hfill \\
$d_k(\genstate{s}{c},\genstate{s'}{c'}) = \sup_{F_k} { |
F_k(\genstate{s}{c}) - F_k(\genstate{s'}{c'})| }$.
\end{definition}

We proceed to show that the uniformities defined by these two
metrics agree.  Unlike the case with discrete time systems the
metrics themselves do not agree: it is the uniformity that is
common to the two of them.

\begin{theorem}\label{maximum} The uniformity induced by $d_k$ coincides with
  the uniformity induced by $m_{\cF}$, the maximum fixed point of $\cF$.
\end{theorem}
\proof{}

We demonstrate that the identity function is a uniformly
continuous isomorphism between $\cG$ equipped with the metrics
$d_k$ and $m_{\cF}$.

Consider the identity function from domain with metric $m_{\cF}$
and range with metric $d_{k}$. A mutual inductive proof shows
that:
\begin{itemize}
\item Every F-function-expression $F$ satisfies:
\[ |F(\genstate{c}{s}) - F(\genstate{c'}{s'})| \leq
m_{{\cF}}(\genstate{c}{s},\genstate{c'}{s'}) \]
\item Every G-function-expression $G$ satisfies
\[ |G(t)(f) - G(t)(g)| \leq 2 \times J(m_{\cF})(f,g)  \]
\end{itemize}
The key case in this proof is the case for the
F-function-expression $\int G(t)$. For this case, the induction on
$G$ and $k \leq \frac{1}{2}$  yields that $k \times G(t)$ is
$1$-Lipschitz for metric $J(m_{\cF})$.  So, by the definition of
the Wasserstein metric, we get the required inductive result for
$\int G(t)$.

This shows that the identity function from domain with metric
$m_{\cF}$ and range with metric $d_{k}$ is uniformly continuous.

We prove the converse below.  We show that $m_{\cF}$
is dominated by $d_k$.  We use the fact that
the closure ordinal of $\cF$ is $\omega$.  Let $ m_0 = \top,
m_{i+1} = \cF(m)$.  We show by induction on $i$ that each $m_i$ is
dominated by $d_k$.  We proceed in the following two steps:

\begin{itemize}
\item
Let $f,g$ be such that $J(m_i)(f,g) > \epsilon $, where
$\epsilon>0$.  We show that there is a G-function expression such
that for some $t$,  $G(t)(g)=0$ and $G(t)(f)> \epsilon $.

If $J(m_i)(f,g)= \epsilon + \gamma$ where $\epsilon, \gamma>0$,
without loss of generality we can assume that there is a $t$ such that:
\[ (\forall t') \ |t-t'|\leq\epsilon + \gamma \Rightarrow\ m_i(f(t),
g(t')) \geq \epsilon + \gamma \]

From finite-variance, we get a partition
\[ t_0 = t - ( \epsilon + \gamma), t_1, t_2, \ldots, t_n = t + \epsilon+ \gamma
\]
such that:
\begin{itemize}
\item Let $g(t_i) = \genstate{s}{c}$.  Then: $(\forall t_i \leq t < t_{i+1})
g(t) = \langle s, \clock{c}- \rate{r_c}|t-t_i| \rangle $.
\item $|t_i - t_{i+1}| < \frac{\gamma}{2}$
\end{itemize}

From the assumption that $m_i$ is dominated by $d_k$, there are
$F$-function expressions $F_1, \ldots, F_n$  such that $|F_j(f(t)) -
F_j(g(t_i))| > \epsilon + \frac{\gamma}{2}$. Using
lemma~\ref{basic}, we have $(\forall t_i \leq t' < t_{i+1})
m_{\cF}(g(t_i),g(t')) < \frac{\gamma}{2}$, and thus we deduce that
$|F_j(f(t)) - F_j(g(t'))| > \epsilon $, for all $t_i \leq t' <
t_{i+1}$.  Using $\min$ and $h \circ$, without loss of generality,
we can assume that $F_j(f(t)) > \epsilon$ and $F_j(g(t_i)) = 0$.
Consider $\min(F_1,\ldots,F_n)$ and $\cL(F)(t)$.   It evaluates to
$0$ on $g$ and to a value $>\epsilon$ on $f$.

\item  Let $\genstate{s}{c}, \genstate{s'}{c'}$ be such that
$W(J(m_i))(\genstate{s}{c}, \genstate{s'}{c'}) = \epsilon+
\gamma$, where $\gamma>0$.  We show that there is a $F$-function
expression such that $F(\genstate{s'}{c'})=0$ and
$F(\genstate{s}{c})> \epsilon $.

Following the proof of lemma~\ref{finite}, we get finite sets of
traces $L_1, L_2$ satisfying $W(J(m_i))(L_1,L_2) > \epsilon + \frac{\gamma}{2}$
and it suffices to prove the result for finite linear combinations.

From the above item, for each pair of traces (one from $L_1$ and
the other from $L_2$), there are G-function expressions,
$G_{ij}$ that are non-zero only on $f_1,\ldots,f_n$ and zero on
$f'_1,\ldots,f'_m$ and which yield arbitrarily close approximations
to the distance between the $f_i, f'_j$ pairs.  The result now follows by
considering $\max(G^p_{ij})$.
\end{itemize}
\qed

\section{Conclusions}

We have given a pseudo-metric analogue of bisimulation for GSMPs.  We have
shown that this really depends on the underlying uniformity and that
quantities of interest are continuous in this metric.  We have given a
coinduction principle and a logical characterization reminiscent of
previous work for weak bisimulation of a discrete time concurrent Markov
chain.

The previous approaches to bisimulation work well for CTMCs, precisely
because of the fact that the distribution is memoryless; at any given
instant the expected duration in a state and the transition probabilities
only depend on the current state of the system, and thus one \emph{can}
define a bisimulation on the state space.  In contrast, the problem of
describing bisimulation for real-time processes that have general
distributions, rather than memoryless distributions, has been vexing.  In
the present work, we have shifted emphasis to the generalized states that
incorporate time and not tried to define a bisimulation on the ordinary
states. Because the generalized states embody the quantitative temporal
information we \emph{have to work metrically}; an attempt to define
bisimulation directly would have fallen afoul of the approximate nature of
the timing information.

If we want to move to continuous state spaces and stochastic hybrid
systems, the whole dynamical formalism has to be different: one can no
longer think of paths as cadlag functions.  We will have to use
stochastic differential equations to describe the systems and the space of
sample paths for the trajectories.  That is a subject for future work and
one that we have been heading towards from the inception of our work on
LMPs~\cite{Blute97,Desharnais02}.

\bibliographystyle{alpha}

\newcommand{\etalchar}[1]{$^{#1}$}

\end{document}